%% file: ms.tex
\pdfoutput=1
\documentclass[sigplan,10pt, authorversion,nonacm]{acmart}
\setcopyright{none}

\usepackage{graphicx}
\usepackage{tabularx}
\usepackage{listings}
\usepackage{todonotes}
\usepackage{booktabs}
\usepackage{enumitem}
\usepackage{microtype}

\usepackage{balance}

\begin{document}


\lstdefinestyle{python}{
  language=Python,
  columns=fullflexible,
  xleftmargin=1em,
  xrightmargin=1em,
  commentstyle=\color{blue},
  basicstyle=\small,
}

\lstdefinestyle{augmented_python}{
  basicstyle=\small,
  frame=single, 
  columns=fullflexible,
  xleftmargin=1em,
  xrightmargin=1em,
  morekeywords={<|cell|>, <|endofcomment|>}, 
  otherkeywords={<|cell|>, <|endofcomment|>}, 
  keywordstyle=\color{blue}
}

\title{Natural Language-Guided Programming}

\author{Geert Heyman}
\email{geert.heyman@nokia-bell-labs.com}
\affiliation{%
  \institution{Nokia Bell Labs}
  \country{Belgium}
}

\author{Rafael Huysegems}
\email{rafael.huysegems@nokia-bell-labs.com}
\affiliation{%
  \institution{Nokia Bell Labs}
  \country{Belgium}
}

\author{Pascal Justen}
\email{pascal.justen@nokia-bell-labs.com}
\affiliation{%
  \institution{Nokia Bell Labs}
  \country{Belgium}
}

\author{Tom Van Cutsem}
\email{tom.van\_cutsem@nokia-bell-labs.com}
\affiliation{%
  \institution{Nokia Bell Labs}
  \country{Belgium}
}

\begin{abstract}
  \input{abstract.tex}
\end{abstract}

\ccsdesc[500]{Software and its engineering~Integrated and visual development environments}
\ccsdesc[500]{Computing methodologies~Natural language processing}

\keywords{\sloppy code completion, code prediction, natural language-guided programming, example-centric programming}

\maketitle

\input{introduction.tex}

\input{vision.tex}

\input{languagemodels.tex}

\input{case_study.tex}

\input{related_work.tex}

\input{research_agenda.tex}

\input{conclusion.tex}

\begin{acks}
We would like to thank our colleagues Frederik Vandeputte, Bart Theeten, Maayan Goldstein, Guillermo Rodriguez-Navas and Cecilia Gonzalez-Alvarez for discussions and their help collecting and labeling the data used in our experiments.
\end{acks}

\input{appendix.tex}

\bibliographystyle{ACM-Reference-Format}
\balance
\bibliography{base}


\end{document}

%% file: abstract.tex
In today's software world with its cornucopia of reusable software libraries, when a programmer is faced with a programming task that they suspect can be completed through the use of a library, they often look for code examples using a search engine and then manually adapt found examples to their specific context of use.
We put forward a vision based on a new breed of developer tools that have the potential to largely automate this process.
The key idea is to adapt code autocompletion tools such that they take into account not only the developer's already-written code but also the \emph{intent} of the task the developer is trying to achieve next, formulated in plain natural language. We call this practice of enriching the code with natural language intent to facilitate its completion \emph{natural language-guided programming}.

To show that this idea is feasible we design, implement and benchmark a tool that solves this problem in the context of a specific domain (data science) and a specific programming language (Python). Central to the tool is the use of language models trained on a large corpus of documented code.
Our initial experiments confirm the feasibility of the idea but also make it clear that we have only scratched the surface of what may become possible in the future. We end the paper with a comprehensive research agenda to stimulate additional research in the budding area of natural language-guided programming.

%% file: introduction.tex
\section{Introduction}
\label{sec:introduction}

In many areas of software development, developers find themselves spoiled with thousands of readily available software libraries (also commonly called modules or packages). The growing practice of building software out of \emph{open source} components further adds to that trend\footnote{According to a 2019 survey ran by Open Source consultancy firm Tidelift up to 93\% of surveyed applications used open source components and up to 70\% of the codebases of surveyed applications consisted of open source code.~\url{https://tidelift.com/subscription/managed-open-source-survey}}, making modern software stacks highly diverse and fast-evolving.

To make this more concrete, consider a data scientist using Python for data analysis.
A trained data scientist familiar with the ecosystem will typically combine a variety of libraries to complete any given job. She might use Python's built-in libraries to download or manipulate raw data files, use the popular \texttt{Pandas} library to manipulate tabular data, use scientific computing packages such as \texttt{NumPy} to manipulate numerical data, use large machine learning frameworks such as \texttt{Scikit-learn} to train predictive models on the data and finally use one of the many popular data visualization libraries such as \texttt{Matplotlib} or \texttt{Plotly} to chart her insights.

If the data scientist is in need of training sophisticated deep neural network models to fit the data, she is spoiled with choice among multiple large and high-quality libraries that will help her do that with just a few lines of code. Two of the most widely used libraries are \texttt{Tensorflow} and \texttt{Pytorch}. Unfortunately for our data scientist, these major machine learning toolkits are constantly tweaking their APIs. In the last three years alone, \texttt{Tensorflow} has received no less than 16 major releases while \texttt{Pytorch} saw 9 major releases\footnote{https://medium.com/analytics-vidhya/pytorch-is-growing-tensorflow-is-not-6986c5e52d6f, retrieved April 2021.}.

\begin{figure*}
    \centering
    \includegraphics[width=0.95\textwidth]{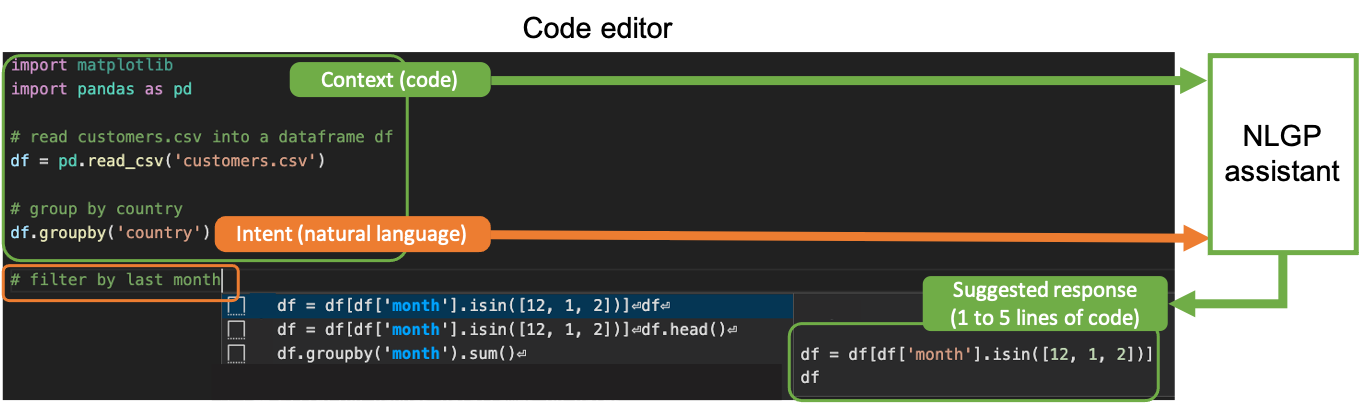}
    \caption{In natural language-guided programming, a programmer formulates tasks using natural language in a specific code context. An NLGP assistant suggests relevant code that matches the task and the context.}
    \label{fig:lgp}
\end{figure*}

Keeping up with the large and ever-changing ``API surface'' of all of these combined library dependencies poses a serious learning challenge to any prospective data scientist. This problem is neither specific to data science nor is it specific to Python. Other domains of software development feature a similarly rich and sprawling software library ecosystem~\cite{xu20reinventing}.

\subsection{Example Embedding to the Rescue?}

Luckily the growing body of APIs is accompanied by growing sources of online documentation. The proliferation of online code examples embedded in library documentation, tutorials or Q\&A websites such as Stack Overflow\footnote{www.stackoverflow.com} has led to a programming phenomenon called \emph{Example Embedding}~\cite{barzilay11example}, also known as \emph{example-centric programming}~\cite{brandt10example}. A developer engages in example embedding when they search for code examples (in local or online repositories), copy-paste the code into their code editor and then adapt the code to fit their specific needs and their specific code context. 

The activity of example embedding is largely done manually, with little dedicated tool support, which can make it time-consuming and error-prone. Indeed, prior empirical studies in software engineering found that up to 35\% of a developer's worktime can be spent on code search~\cite{xia17developers}. There is also evidence that suggests that code found in documentation or online knowledge bases is rarely immediately usable. For example, one study of Stack Overflow found that a mere 25.61\% of Python code snippets could be readily executed, and the figures were even worse for other languages~\cite{yang2016query}. Even when the developer does find a high-quality code example, they must still edit the code to fit their specific context, e.g. by renaming variables or deleting superfluous statements. These edits are an opportunity for bugs to creep into the developer's codebase.

Given these observations, we conjecture that automating the activity of example embedding has the potential to positively affect both developer productivity as well as code quality.

\subsection{Automating Example Embedding}

We envision that this common practice of example embedding will become more and more automated through new tools that leverage advances in machine learning and natural language processing. We will refer to the practice of using tools to automate the example embedding process as \emph{natural language-guided programming} (abbreviated NLGP). In a coding environment that supports natural language-guided programming, when the programmer is faced with a task that they believe can be solved using a library, they simply state their intent using natural language in-line in the code and an NLGP tool suggests code that 1) addresses the task and 2) fits the context, choosing variable and function names that match the already existing code. 
We will refer to such a developer tool as an \emph{NLGP assistant}. The diagram in Figure~\ref{fig:lgp} describes the basic NLGP workflow.

Much like refactoring assistants in modern IDEs now help software developers more quickly and reliably apply refactorings to their codebase, we envision that natural language-guided programming tools will help developers more quickly and reliably perform example embedding. The goal of an NLGP assistant is to help programmers write idiomatic code for tasks that can be solved using available libraries.

\subsection{Paper Contributions}

\begin{itemize}
    \item We demonstrate our vision of natural language-guided programming in the domain of data science and machine learning using Python libraries (Section~\ref{sec:case_study}).
    \item We contribute the design and implementation of an NLGP assistant for this domain. The core of our approach is based on language models (Section~\ref{sec:langmodels}) which require training on a large corpus of documented code. We develop three language model variants to study the impact of natural language intent on prediction quality together with a benchmark and a user evaluation (Section~\ref{sec:implementation}).\footnote{The models are shared on \url{https://huggingface.co/Nokia}, and the benchmark and user annotations can be downloaded from \url{https://zenodo.org/record/5384768\#.YTDsN9MzZUJ}.}
    \item We articulate a Research Agenda for natural language-guided programming: what are key open research questions that need to be addressed to move this field forward? (Section~\ref{sec:research_agenda}).
\end{itemize}

%% file: vision.tex
\section{Case Study: NLGP for Data Science and Machine Learning in Python}
\label{sec:case_study}

In the introduction we described how a data scientist would typically use a multitude of Python libraries to do their job. Let us now walk through a concrete experience of what it would be like to perform data analysis using natural language-guided programming.

Before we get started, we give a brief background on the typical programming environment used by data scientists.

\subsection{Background: the Python Data Science Stack}

In recent years Python has become the language of choice for an increasing number of data scientists, data engineers, and machine learning researchers.\footnote{https://towardsdatascience.com/top-programming-languages-for-data-science-in-2020-3425d756e2a7, retrieved April 2021} As mentioned in the introduction, one reason for this is Python's large ecosystem of scientific computing libraries, sometimes called the ``Python data science stack'' or simply the ``data stack''.\footnote{https://hub.packtpub.com/python-data-stack/, retrieved April 2021.} Table~\ref{tab:datastack} lists key projects in this stack.

\begin{table}[h]
    \caption{Key projects in the Python Data Stack}
    \label{tab:datastack}
    \centering
    \begin{tabular}{p{0.2\columnwidth} p{0.70\columnwidth}}
        \hline
        \textbf{Project} & \textbf{Purpose} \\
        \hline
        NumPy & N-dimensional arrays and extensive math operations. \\
        \hline
        SciPy & Advanced math (solvers, optimizers). \\
        \hline
        Pandas & Rich data manipulation for tabular data. \\
        \hline
        Matplotlib & 2D data plotting. \\
        \hline
        Scikit-learn & Comprehensive machine learning toolkit. \\
        \hline
        Jupyter & Interactive notebooks with text, code, math and graphics. \\
        \hline
    \end{tabular}

\end{table}

A Jupyter notebook is an interactive coding environment composed of \emph{cells}. The two most commonly used cells are code cells and markdown cells. Markdown cells contain simple markup text and can be rendered to a variety of formats. Code cells may contain code in one of the languages supported by the Jupyter protocol. Here we will focus only on Python code.

A code cell can be executed after which the result of the code is inserted as output in the notebook. Jupyter has built-in support to render certain program values as rich media (such as tables or graphics). This allows a data scientist to easily inspect the intermediate output of their data transformations. This interactive style of programming aligns well with NLGP because it allows for programmers to rapidly test and explore auto-generated code by executing it.

\subsection{Natural Language-Guided Programming in Jupyter}

Let us put ourselves in the position of Dana the data scientist. Dana is tasked with analyzing stock market prices stored in a comma-separated value (CSV) file.

\begin{figure}
    \centering
    \includegraphics[width=0.9\columnwidth]{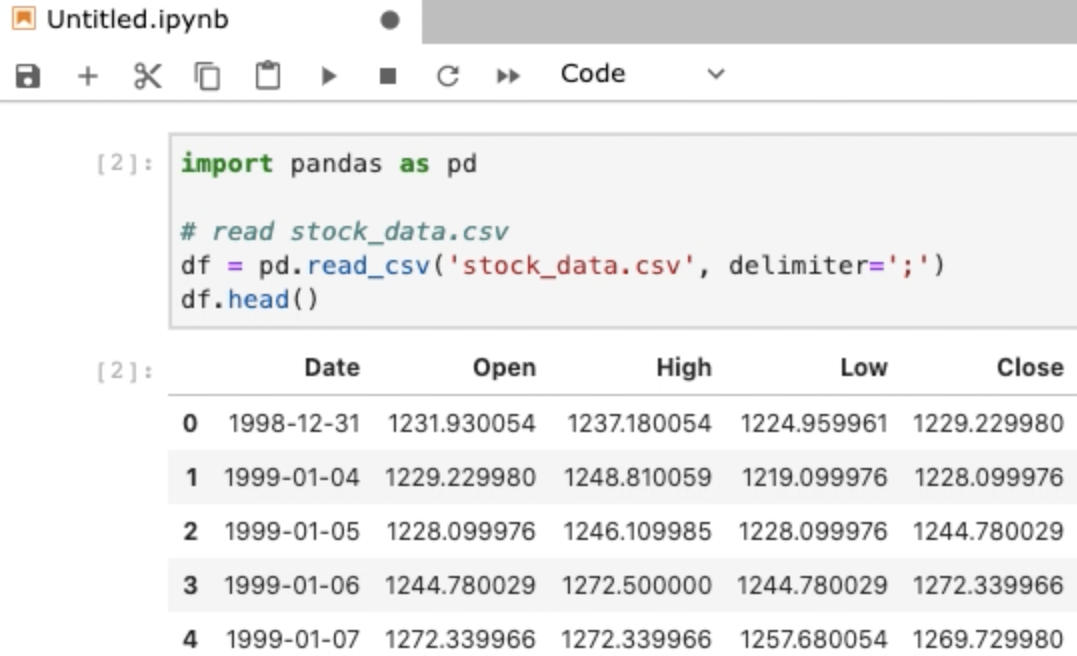}
    \caption{Jupyter notebook with the result of executing the code cell after inserting the NLGP assistant's suggested code.}
    \label{fig:jupyterscreen1}
      \vspace*{-1em}
\end{figure}

Dana knows that Pandas is the go-to library to manipulate tabular data like this, so she starts by importing the library:

\begin{lstlisting}[frame=single, style=python]
import pandas as pd
\end{lstlisting}

\sloppy The Pandas library offers a data structure called a ``dataframe'' to manipulate
tabular data. Dana now needs to read the CSV file into memory and convert it into such a dataframe. She knows that Pandas has an API for this but forgot the details. Rather than looking up the right API call in the documentation or searching the Web for an example, using an NLGP assistant Dana can simply write her intent as a single-line comment in the code:

\begin{lstlisting}[frame=single, style=python]
import pandas as pd
# read stock_data.csv
\end{lstlisting}

Dana then triggers her NLGP assistant using a hotkey, which looks at her code and her intent and produces the following suggestion:

\begin{lstlisting}[frame=single, style=python]
df = pd.read_csv('stock_data.csv', delimiter=',')
\end{lstlisting}
    
That suggestion looks relevant so Dana hits 'enter' to insert the line into the code cell:

\begin{lstlisting}[frame=single, style=python]
import pandas as pd
# read stock_data.csv
df = pd.read_csv('stock_data.csv', delimiter=',')
\end{lstlisting}

Once the suggested code is merged, Dana is free to modify it. Perhaps the CSV file used a ';' separator rather than a ',' separator. Dana can easily update the suggested parameter. At this point, Dana can run the code to inspect the contents of \verb|df| and verify that the data was parsed correctly. Figure~\ref{fig:jupyterscreen1} shows the output of the above
code cell in a Jupyter notebook environment. The call to \verb|df.head()| was added to visualize the first five rows of the dataframe.

Now Dana would like to select only a subset of the columns. Again, rather than looking up how to do this in the documentation, she can state her intent as a comment:

\begin{lstlisting}[frame=single, style=python]
# select columns 'Date', 'Open' and 'Close'
\end{lstlisting} 

\pagebreak
The NLGP assistant suggests the following lines of code:

\begin{lstlisting}[frame=single, style=python]
df = df[['Date', 'Open', 'High', 'Low', 'Close']]
df.head()
\end{lstlisting}

Even though the suggestion contains some spurious columns, it looks mostly right to Dana, so she incorporates the suggestion and makes a few edits:

\begin{lstlisting}[frame=single, style=python]
# select columns 'Date', 'Open' and 'Close'
df = df[['Date', 'Open', 'Close']]
df.head()
\end{lstlisting} 

Dana can continue to write code in this way by stating her intent upfront and accepting relevant code suggestions into her notebook:

\begin{lstlisting}[frame=single, style=python]
# convert 'Date' column to datetime value
df['Date'] = pd.to_datetime(df['Date'])

# add a new column called 'Month'
df['Month'] = df['Date'].dt.month

# group by Month and compute the average
df.groupby('Month').mean()

# chart the data
df.plot()
\end{lstlisting} 

In the above code, the comments in blue indicate the programmer's ``intent''.
The NLGP assistant is invoked after entering the intent. The code following each intent is generated by the NLGP assistant.
The code suggestions presented above are actual code predictions made by the tool introduced in Section~\ref{sec:implementation}.

The above NLGP session focused on assisting Dana with the \texttt{Pandas} library only, but we expect an NLGP assistant to be able to offer suggestions for tasks requiring other libraries through the same unified interface. For example, Dana might also use the popular \texttt{Scikit-learn} library for machine learning tasks. Example intents that she might use to describe such tasks include:

\begin{itemize}
  \item ``split the data in a training and a validation set''
  \item ``cluster the data using K-means''
  \item ``plot a confusion matrix of the classifier''
\end{itemize}

The user stating their intent is an integral part of natural language-guided programming. However, we make no assumptions on how that intent is communicated to the NLGP assistant. In the NLGP session described above, the developer states their intent as an in-line comment in the code. Other NLGP assistants may offer a text field separate from the code to guide the code autocompletion. One potential benefit of inlining the intent in the code is that it self-documents the interaction with the tool, forming a potential source of future data to better learn the translation from intent to code.

In the next section, we introduce language models as a general technique for predicting text or code. Subsequently, we show how an NLGP assistant for Python can be built using this technique.

%% file: languagemodels.tex
\section{Background: Language Models}
\label{sec:langmodels}

Language models are statistical models that estimate the probability of a given sequence of tokens, such as words in an English-language text with respect to a reference corpus. This is a very general problem with many applications, including text autocompletion, speech recognition and machine translation~\cite{jurafsky2008speech}. In the past decade, advances in the field of deep learning, such as the Transformer model architecture~\cite{vaswani17attention}, have significantly improved the effectiveness of language models on practical applications.
One particularly successful set of language models based on Transformers is the GPT (Generative Pretrained Transformer) model family including GPT-2~\cite{radford2019language} and its successor GPT-3~\cite{Brown2020}. GPT models have been trained on large volumes of text from public Web pages. Their capability to generate seemingly human-written text has received widespread attention both within and outside the research community.\footnote{See e.g. this New York Times article dd. July 29, 2020: \url{https://www.nytimes.com/2020/07/29/opinion/gpt-3-ai-automation.html}}

Hindle \emph{et al.}~\cite{10.5555/2337223.2337322} were the first to formally observe that code, like natural language, is repetitive and predictable and that language models can also be used to create effective statistical models of source code, paving the way towards new kinds of code completion tools. Hindle \emph{et al.}'s work was based on (adapted versions of) n-gram models and in recent years there has been an ongoing debate about what type of language models (based on n-grams or deep learning) is best for modeling source code~\cite{hellendoorn2017deep, karampatsis2019maybe}. In recent years, in line with what has been observed in NLP in general, language models based on deep learning such as Transformers have been shown to achieve production-quality levels of code completion, with companies offering code completion products based on this technology.\footnote{See  Tabnine blog dd. July 15, 2019 \url{https://web.archive.org/web/20201204055827if_/https://www.tabnine.com/blog/deep}, archived Dec. 4, 2020 and GitHub CoPilot, \url{https://copilot.github.com/}, retrieved Aug. 3, 2021.}

We now review how language models can be used to generate sequences that fit a given context. A language model estimates the probability distribution $P(X_{n} | X_{1}, ..., X_{n-2},  X_{n-1})$ of the n$^{th}$ token $X_n$ given the previous tokens in the sequence.  With this conditional probability distribution, we can predict the most likely token in a given context. By adding the predicted token to the context, we can iteratively expand the prediction to form sequences of arbitrary length. Generating the most likely sequence is intractable\footnote{The space and time complexity for generating the most likely sequence scales exponentially: $O(|V|^L)$, where $|V|$ is the vocabulary size and $L$ is the sequence length.} so in practice approximate algorithms such as beam search~\cite{graves2012sequence} are used to explore the search space.

To make the problem tractable, most language models make the simplifying assumption that $X_{n}$ only depends on a window with the $C$ previous tokens: $P(X_{n} | X_{1}, ..., X_{n-2},  X_{n-1}) = P(X_{n} |  X_{n-C}, ..., X_{n-2},  X_{n-1})$. For instance, GPT-2 language models can process a maximum of 1024 tokens at a time, which means that $C$ can be at most 1023.

Language models can be applied to different types of token sequences: tokens can correspond to words, subword units, or individual characters. When applying language models to source code, where the number of unique identifiers tends to be large~\cite{10.5555/2337223.2337322}, subword units are desirable.
When we discuss training of language models on code in this work, we assume the use of byte-pair encoding (BPE)~\cite{gage1994new, sennrich2016neural} as used in GPT-2. BPE is a compression algorithm for splitting words into subword tokens such that the most frequent (sub)words are tokenized as a single token. For example, applying the GPT-2 tokenizer to the code string `\verb|b = np.zeros(10)|' would result in the following subword units: `\verb|b|', `\verb|␣=|', `\verb|␣np|', `\verb|.|' , `\verb|zer|', `\verb|os|', `\verb|(|', `\verb|10|' and `\verb|)|'.

In the next section, we describe how an effective NLGP assistant can be built based on the GPT-2 model.

%% file: case_study.tex
\section{Building an NLGP Assistant using Language Models}
\label{sec:implementation}

To study the feasibility of NLGP we build and evaluate a prototype of an NLGP assistant for Python. Our NLGP assistant uses a language model to autocomplete code cells based on both existing code in the cell, as well as the developer's intent, specified as a comment (as introduced in Section~\ref{sec:case_study}). The language model is trained on a collection of preprocessed Jupyter notebooks (details of our dataset are covered in Section~\ref{sec:jupyterdata}).

We first introduce three strategies for preprocessing the data, leading to three distinct language models. Next, we cover more details on how we prepare the data and train the models. Finally, we report on an initial user study to evaluate the quality of the models' code predictions.

\subsection{Language Models for NLGP}

The starting point for all of the language models trained in this paper is the GPT-2 Medium model checkpoint released by OpenAI~\cite{radford2019language}. The model checkpoint was pretrained by OpenAI on general-purpose English-language text crawled from the Web. From preliminary experiments we concluded that starting from a pretrained model gave significantly better results than starting from an equivalent randomly initialized transformer model that was not pretrained on text.

To train a language model to be able to autocomplete code based on existing code and a natural language intent, we need relevant training data. The challenge here lies in finding a sufficiently large amount of code that is self-documented with the developer's intent. Given that there exists no sufficiently large dataset of Python code that is explicitly annotated with the developer's intent using natural language\footnote{We survey relevant datasets in Section~\ref{sec:related_work}.}, we need creative ways to teach the language model how to associate natural language intent with code. One assumption is to rely on textual comments in the code. We consider three distinct ways to use comments in code to train language models:

\textbf{No Comments} The \texttt{no comments} model is trained on a dataset where all original comments are stripped from the training data. This model serves as a baseline and will allow us to quantify how important it is to consider natural language intent in addition to pure code.
   
\textbf{Docstring Comments} The \texttt{docstring} model is trained on a dataset where we also first strip all comments from the training data. However, here we annotate a selection of call sites with synthetic comments. These comments contain a summary of the called method's or function's docstring. The intuition is that a docstring typically contains a short one-sentence description of the intent of the function or method. We describe this procedure in detail in Section~\ref{sec:data_augmentation}.
  
By annotating the call site with the docstring, we hope to teach the model to associate code context preceding the call with keywords from the docstring and the subsequent method or function call. This setup is meant to assess the feasibility of NLGP models in domains where code is not documented with relevant comments.

\textbf{Natural Comments} The \texttt{natural} model is trained on comments interleaved with code as they naturally occur in Jupyter notebooks. This includes text in markdown cells as well as in-line comments in code cells. In this dataset no call sites are annotated with docstrings.

\subsection{Jupyter Notebook Dataset}
\label{sec:jupyterdata}

%
%

Jupyter notebooks are a mix of executable code and descriptive text. This makes them an interesting source for collecting training and evaluation data for an NLGP assistant.
To construct a dataset, we searched GitHub for all projects that contain at least one Jupyter notebook, have a permissive license and received at least one star.
Next, we apply a heuristic to filter out project forks: when multiple projects have the same name, only the project with the most stars is retained. We then download all notebooks in the project and convert them to \texttt{.py} source files using the \texttt{nbconvert} tool.\footnote{\url{https://nbconvert.readthedocs.io/en/latest/}}$^{,}$\footnote{We skipped notebooks containing code written in languages other than Python (e.g. Julia, R), as well as notebooks under .ipynb\_checkpoint/ folders.} This tool converts any non-code cells into inline comments. We parse the \texttt{.py} files using a Python3 parser and reject any files that contain parse errors. The resulting files are split 90/10 across a training and evaluation set. We ensure that notebooks that belong to the same GitHub project end up in the same split. In this way, we obtain 297,845 and 32,967 \texttt{.py} files for training and evaluation purposes respectively.

Each \texttt{.py} file in the training split was further preprocessed and cleaned using following heuristics:
\begin{itemize}
\item Any markdown content before the first code cell delimiter is removed;
\item Comments that were inserted by \texttt{nbconvert} to delimit code cells (\# In [], \# In [1] , \# In [2], etc. ) are replaced by a special \texttt{<|cell|>} token;
\item Comments are separated from the subsequent code by a special \texttt{<|endof\-comment|>} token (more details below);
\item Multi-line comments are truncated to a maximum of two lines;
\item Markdown header symbols, which are inserted by the \texttt{nbconvert} tool, are stripped (e.g., \texttt{\# \#\# some title} is converted to  \texttt{\# some title});
\item Non-English comments are stripped. We used the \texttt{cld3} tool\footnote{\url{https://github.com/google/cld3}} to automatically detect the language;
\item Empty cells and empty comments are removed.
\item Spaces are replaced by special whitespace tokens (e.g., '\texttt{    } ' is replaced by a single '\texttt{<|4space|>}' token).
\end{itemize}


\subsection{Language Model Setup for Intent-Guided Code Prediction}

To use a language model to generate predictions in an NLGP context, two issues remain: 1) What is the stopping criterium (when has the model predicted enough code to address the intent)?; 2) How to force the model to predict source code instead of autocompleting the inline comment with more natural language? If the model were to autocomplete the intent, it may inadvertently change its meaning, which is undesirable.

To address these challenges, we introduce additional symbols <|endofcomment|> and <|cell|> to encode structural information, as illustrated in the following example:

\noindent \begin{minipage}{\linewidth}
\begin{lstlisting}[style=augmented_python]
  ...
  # Choose number of features automatically
  # use RFECV to select features <|endofcomment|>
  rfe = RFECV(random_forest, n_jobs=-1, step=1)
  rfe.fit(X_train, y_train)
  
  feature_scores['RFECV'] = X.shape[1] - \
      rfe.ranking_.astype(float).reshape(-1, 1)
  <|cell|>
  # output number of features <|endofcomment|>
  print("#features=", np.sum(rfe.support_))
  <|cell|>
  \end{lstlisting} 
\end{minipage}

An \texttt{<|endofcomment|>} token is inserted after every inline comment that is followed by source code. That is, for multiple successive inline comment lines, we only insert the token after the last comment line. At prediction time, we append this symbol to the end of the user intent to prompt the model to predict source code rather than to autocomplete the comment. A \texttt{<|cell|>} token is inserted at the end of every Jupyter notebook code cell. At prediction time, no more tokens are predicted after the model has predicted a \texttt{<|cell|>} token. We found that this simple heuristic works well in practice, but there is room to experiment with more sophisticated stopping criteria in future work.

As a final step, we concatenate all preprocessed \texttt{.py} files into a single training file using the \texttt{<|endoftext|>} symbol to encode the original file boundaries.

We generate predictions using beam search with a beamwidth of 3, where the prediction of the \texttt{<|cell|>} token signals that a beam hypothesis is complete. We enforce that the model predicts between 10 and 150 tokens by setting the probability of the stopping token to zero for the first 10 tokens and by stopping the beam search procedure after the beam hypotheses are 150 tokens long. The maximum context length is set to 700 tokens.

We made a slight adjustment to the GPT-2 model and the GPT-2 tokenizer to ensure that our special tokens (\texttt{<|4space|>}, \texttt{<|endoftext|>} , \texttt{<|endof\-comment|>, etc.}) are tokenized as a single token and are encoded with their own set of (embedding) parameters that are initialized at random and trained from scratch. We use the transformers library~\cite{DBLP:journals/corr/abs-1910-03771} to make these changes. 

In Section~\ref{sec:benchmark}, we describe how we used the evaluation split to create a labeled test set.

\subsection{Docstring Comment Injection}
\label{sec:data_augmentation}

The \texttt{docstring} model is trained on a synthetic dataset where all naturally occurring comments in the training data are first removed, after which a random sample of call sites is instrumented with new comments taken from docstrings. 
More specifically, when a call is made to a documented library API, an additional inline comment is added to the code, describing the purpose of the call.
The goal is to augment the source code with comments that capture the intent of the calls using short natural language statements.

For example, given the following snippet of Python code:

\begin{lstlisting}[frame=single, style=python]
from sklearn.cluster import KMeans
k = KMeans()
k.fit(Xtrain)
y = k.predict(Xtest)
\end{lstlisting}

The goal is to transform it into:

\begin{figure*}[t]
  \centering
  \includegraphics[width=\textwidth]{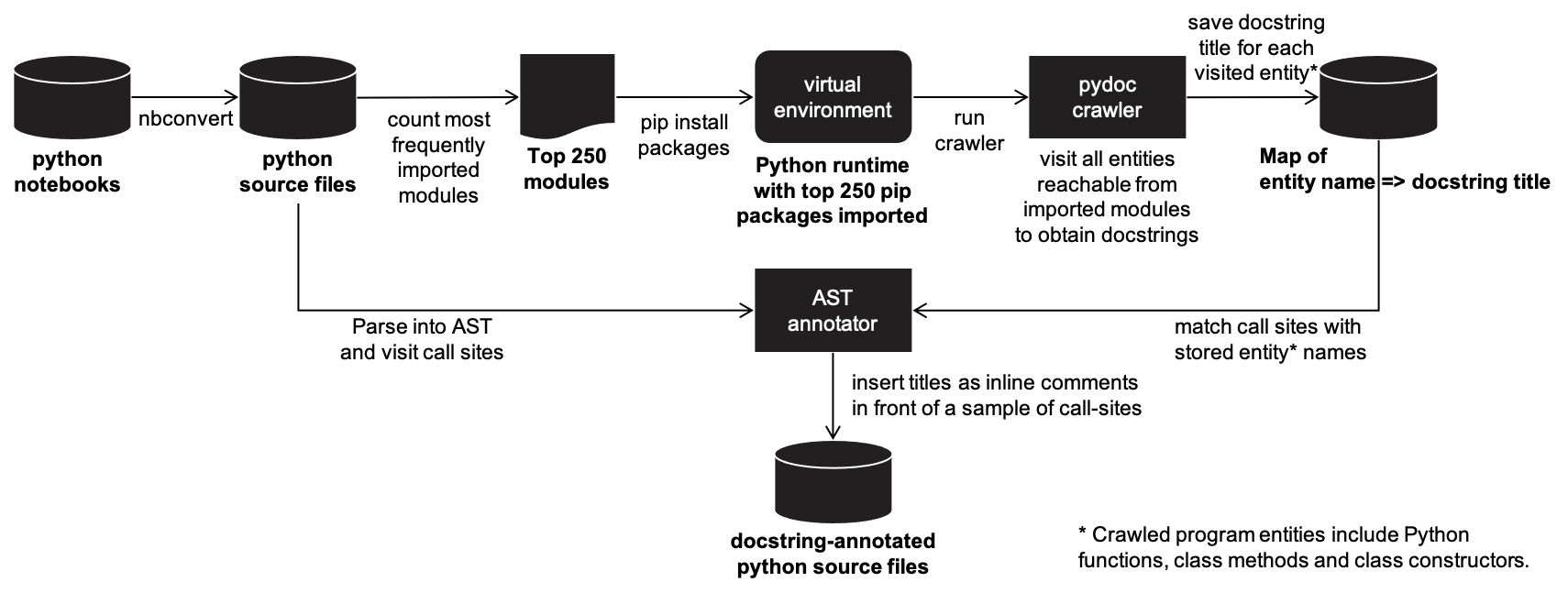}
  \caption{High-level process flow to inject Python docstrings into code.}
  \label{fig:pyannotate}
\end{figure*}

\begin{lstlisting}[frame=single, style=python]
from sklearn.cluster import KMeans
# K-Means clustering
k = KMeans()
# Compute k-means clustering
k.fit(Xtrain)
# Predict closest cluster for each sample
y = k.predict(Xtest)
\end{lstlisting}

Figure~\ref{fig:pyannotate} depicts the high-level process that we followed to implement this transformation. The first objective is to create a mapping from the names of callable program entities (functions, methods, constructors) to their docstrings:

\begin{enumerate}
\item From the python source files in the training set, the root module names are extracted and counted. The 250 most frequently used, non-standard Python library root module names are kept.
\item A blank virtual environment is created in which packages, together with their package dependencies are installed using the `pip` command~\cite{pip}. For most packages, pip is able to install via root module name (e.g. \texttt{numpy}, \texttt{sklearn}, etc). Only a few need an explicit module-package name mapping.
\item Using a custom crawler program all installed packages and standard python libraries/packages are recursively scanned to find all callable program entities. For each harvested entity, the fully qualified pathname (FQPN) and the first sentence from the associated docstring are automatically extracted and stored in the mapping table.
\item We obtain a mapping from FQPN to docstring titles for 64.6\% of the visited callable entities. Entities without an associated docstring are ignored and not recorded in the mapping.
\end{enumerate}

\begin{table}[h!]
\centering
\caption{Example entity-docstring mappings}
\label{tab:mapping}
\resizebox{\columnwidth}{!}{
\begin{tabular}{|l|l|}
\hline
Fully qualified path name & Docstring title\\
\hline
\texttt{sklearn.cluster.KMeans()} & 'K-Means clustering' \\
\texttt{sklearn.cluster.KMeans().predict()} & 'Predict closest cluster each sample ...'\\
\hline
\end{tabular}
}

\end{table}

Table~\ref{tab:mapping} lists a short fragment of the entity-docstring mapping for two entities from the \texttt{sklearn} library. In a second phase, we parse the source files in the dataset and visit all call sites using an AST walker. For each call site, we try to resolve the call to a named entity, e.g. the call \verb|k.predict()| would resolve to \verb|sklearn.cluster.KMeans().predict()|. Because of Python's dynamic typing, we are only able to resolve a subset of calls using a basic program flow analysis. Still, this allows us to resolve 51.3\% of call sites to one of the entity names stored in the mapping.

In a final phase, the AST annotator chooses a random sample of resolved call sites and then inserts the associated docstring title in front of the call. The docstring is always inserted on the previous line of the statement enclosing the visited call site. In our experiments we chose a sampling rate of 20\%. A deeper study of the effect of the sampling rate on the prediction quality is left as future work.

\subsection{Creating an NLGP Benchmark}\label{sec:benchmark}

To assess the prediction quality of an NLGP assistant we need a good benchmark. As no such benchmark exists, we set out to create our own.

A benchmark for NLGP requires realistic scenarios (test cases) where an NLGP assistant needs to complete the code based on a natural language intent. Each test case is a triplet $c$/$i$/$t$ containing a 
code context $c$; a natural language intent $i$, provided in the form of an inline comment; 
and a target code snippet $t$, a reference code snippet that addresses the intent and is a natural completion of the code context $c$.

We created a benchmark containing such triplets in two stages. In a first \emph{generation} stage we automatically mine candidate triplets from the Jupyter notebook dataset. In a second \emph{curation} stage we filter remaining candidates based on human review.

\textbf{Generation Stage} To create realistic and unbiased $c$/$i$/$t$ triplets, we chose to mine triplets from our Jupyter notebook dataset. More specifically, we sample candidate test cases only from source files that were set aside for evaluation (i.e. \emph{not} occurring in the training set):

\begin{enumerate}
\item We scan the source files for lines that only contain an inline comment and whitespace.
\item Next, we filter out non-English comments and comments that are longer than 10 tokens. This cut-off was informed by studying the query lengths of the user study done by Xu \emph{et. al}~\cite{xu2021inide}:  over 97\% user queries consisted of 10 tokens or less.
\item From the remaining comments, we then sample at random and create a set of candidate test cases $c_c$/$i_c$/$t_c$:
\begin{itemize}
\item The candidate context $c_c$ is extracted from the start of the comment's source file up to the line of the comment.
\item The candidate intent $i_c$ is set to the sampled comment including any leading whitespace.
\item The candidate target code $t_c$ is set to all the code (excluding comments) that follows the comment.
\end{itemize}
\item We filter out candidates that overlap with code in the training set. Specifically, we concatenate the last three non-empty lines in the candidate context with the candidate intent and check if the resulting piece of code occurs in the training dataset. If an exact match is found, the candidate is dropped.
\end{enumerate}

\textbf{Curation Stage} Mined candidate test cases $c_c$/$i_c$/$t_c$ were reviewed by human annotators and refined into representative test cases $c$/$i$/$t$:

\begin{enumerate}
  \item We generate three non-overlapping batches of candidate test cases. Each batch contained 200 distinct cases.
  \item The 3 batches were assigned for review to 9 human reviewers. Each batch was assigned for review to a group of 3 reviewers. As such, a total of 600 candidate test cases were reviewed, each case receiving 3 reviews.
  \item Annotators were asked to decide i) whether the candidate test case is relevant, ii) were allowed to slightly rephrase the candidate intent (e.g. rephrasing a comment in the code like ``and now let's plot the data'' to a more succinct intent like ``plot the data''), and iii) were requested to mark in the candidate target code $t_c$ which specific lines of code $t$ best addressed the intent.  Appendix~\ref{ap:annotation-process} provides further details about the annotation process, including the detailed guidelines that were given to the annotators, a screenshot of the annotation interface, and statistics about the inter-annotator agreement.
  \item When 2 out of 3 reviewers judged that a candidate test case is relevant and the difference between their respective target code selections $t_c$ was not more than 2 lines, the test case was added to the benchmark. 201 out of 600 code snippets were selected in this way.
  \item We postprocess the resulting test cases such that: 
	\begin{itemize}
	\item All the code before the first line of target code is moved to the context
	\item Import statements in the context that are only required for the target code are moved to the target code because it is unrealistic to assume a user will have written such import statements before issuing a query
	\item All comments in the target code (if any) are stripped
	\end{itemize}	  
\end{enumerate}

After going through the curation stage we end up with a benchmark of 201 representative test cases $c$/$i$/$t$ that we can now use to validate the quality of code predictions made by the models. Table \ref{tab:summary_stats} displays some key statistics about the benchmark.

\begin{table}[]
\caption{Summary statistics for the NLGP benchmark. LoC stands for `lines of code".}
\label{tab:summary_stats}
\begin{tabular}{@{}ll@{}}

\toprule
number of samples           & 201  \\
average LoC context         & 268  \\
average LoC target code     & 2.45 \\
average \# tokens in intent & 5.39 \\ \bottomrule
\end{tabular}

\end{table}
\subsection{Evaluation}

We now assess the performance of our language models on the NLGP benchmark. Recall that we trained models on three distinct datasets:

\begin{description}
\item[No comments] A model trained on only code, with comments stripped out. This model serves as a baseline to measure the importance of natural language intent;
\item[Docstring] A model trained on code augmented with injected docstring comments on 20\% of calls to APIs from libraries documented with pydoc docstrings.
\item[Natural] A model trained on code including all the comments that occur naturally in the code.
\end{description}

For each model, we create a code prediction $p$ for each $c$/$i$/$t$ triplet in our benchmark. We provide concrete examples of $c$/$i$/$t$/$p$ cases in Appendix~\ref{ap:examples}. The average prediction latency on an 11GB GeForce GTX 1080 Ti GPU was 1.87 seconds\footnote{
In work following the reported experiments, we used the ONNX runtime framework (\url{https://www.onnxruntime.ai/}) to bring down the average prediction latency of these models down to 0.3 seconds, which is sufficiently fast to enable interactive code predictions within an IDE.}.

To assess how well the generated code prediction $p$ compares to the reference code $t$, we set up a human evaluation study. We first introduce the study, then discuss how the human evaluation results correlate with standard text comparison metrics such as BLEU~\cite{10.3115/1073083.1073135}.

\subsubsection{Human Evaluation Study: Setup}

Using the prepared $c$/$i$/$t$/$p$ entries, we ran a small human evaluation study where the first three authors of the paper manually scored the code predictions of each model across four dimensions: usefulness, coverage, precision, and compatibility. Specifically, for the predictions of each model on 100 test cases in our benchmark, each participant rates the following statements on a 4-point scale (Strongly disagree, Disagree, Agree, Strongly agree):

\begin{description}
    \item[Usefulness] The predicted code is helpful for implementing the given intent;
    \item[Coverage] The predicted code completely covers the intent;
    \item[Precision] The predicted code mostly contains code that is relevant to the intent;
    \item[Compatibility] The predicted code is compatible with the code context (e.g., it reuses variable names from the context if appropriate)
\end{description}

To avoid that the annotators have a bias to a certain model, the predictions were presented in random order and the annotation interface did not display the model name.

\input{user_study_results}

\input{metrics}

%% file: user_study_results.tex
\subsubsection{Human evaluation study: results}

\begin{figure*}
    \includegraphics[width=\textwidth]{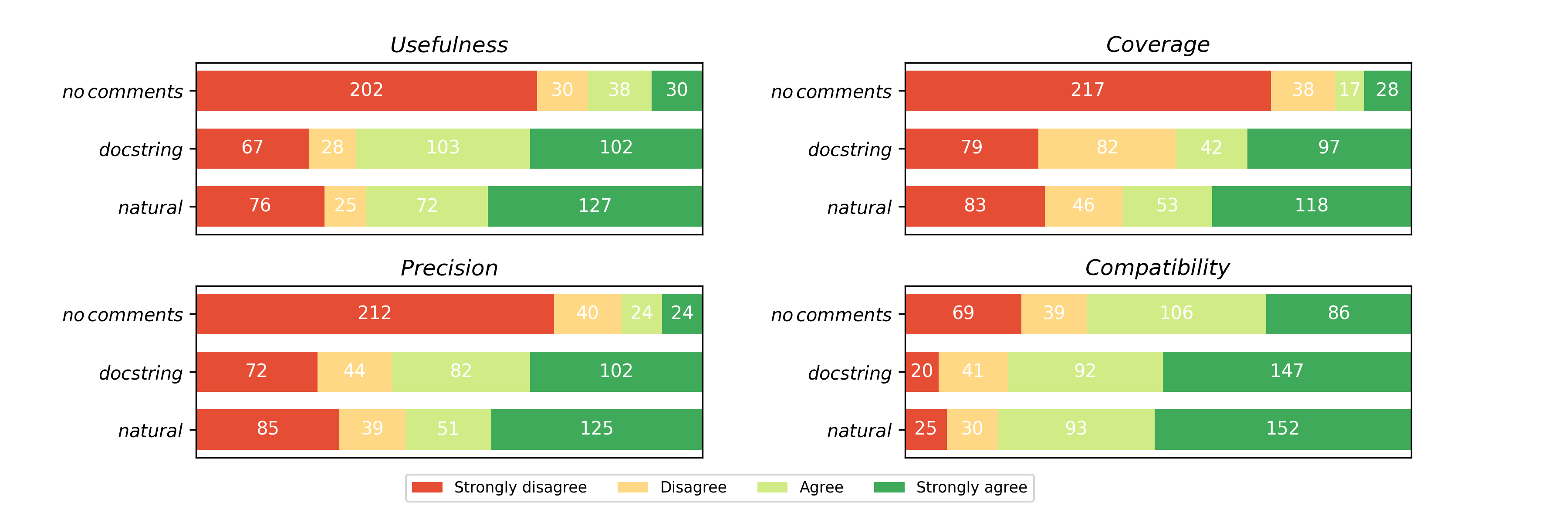}
    \caption{Distributions of the answers in the prediction quality survey: three data science experts each assess model predictions for 100 test cases w.r.t. claims about their usefulness, coverage of the intent, precision, and compatibility with the existing source code.}
    \label{fig:result_user_study}
\end{figure*}

Figure~\ref{fig:result_user_study} reports the answer distributions to the survey questions for each model.
The results indicate that the models trained on comments significantly outperform the \texttt{no comments} model. 
As expected, it is more difficult for models to guess the programmer's intent from only undocumented code.

Both the \texttt{docstring} and \texttt{natural} models exhibit decent performance with similar overall scores even though the \texttt{natural} model results in a better intent coverage. This difference can be attributed to the fact that the original inline comments are more diverse than docstring titles supporting the fact that the \texttt{natural} model can translate a more diverse set of intents. The relatively small gap between the two does indicate that even in domains where code is not heavily documented the NLGP approach is feasible when a procedure similar to our docstring-injection is feasible. The models score particularly well w.r.t. compatibility, implying that the models can generate code predictions customized to the code context.

Finally, the usefulness scores of both the \texttt{docstring} and \texttt{natural} models reflect that the majority of their predictions are considered useful. These results support the feasibility of natural language-guided programming and suggest that our NLGP assistant prototype may already be helpful to support data scientists in practice.


%% file: metrics.tex
\subsubsection{Metrics}
\label{sec:metrics}

Running human evaluation studies to validate code prediction models is time-consuming. 
For this reason, it is desirable to have good metrics that can automatically score predicted code.
One way to accomplish this is to compare the predicted code with the reference target code (also called the ``ground truth'' code).

Previous work to measure the quality of predicted code~\cite{iyer-etal-2018-mapping,yin2018tranx,DeepApi,xu-etal-2020-incorporating,clement2020pymt5} mostly treats the code as plain text and uses the Bilingual evaluation under study (BLEU) score~\cite{10.3115/1073083.1073135}. BLEU is a standard metric in natural language text translation. The BLEU score is based on the ratio of common n-grams found in the prediction and the reference text (the target code).

Intersect-over-Union (IoU), also known as Jaccard distance, between sets of tokens derived from code fragments has also been used to evaluate code prediction tools. For example, Murali \emph{et al.} compute Jaccard distance between sets of API calls occurring in the predicted and the target code~\cite{murali2018neural}.

We computed both BLEU and IoU metrics for the code predictions in our benchmark and correlated them with the usefulness scores that were assigned in the human evaluation study. Our goal here is to measure how well these metrics can act as a proxy for the average usefulness score assigned by human experts.

Before applying the metrics, the predicted code and the target code are tokenized using the tree-sitter library~\cite{treesitter}.  

The tokenization is illustrated in the following example. 

Original code:
\begin{lstlisting}[frame=single, style=python]
from matplotlib import pyplot as plt
plt.hist(means_100)
plt.show()
\end{lstlisting}

Resulting tokens used to compute the metrics:
\begin{lstlisting}[frame=single, style=python]
from matplotlib import pyplot as plt plt . hist 
( means_100 ) plt . show ( )
\end{lstlisting}

\begin{table}[]
\caption{Average BLEU and IoU score on the benchmark, along with their correlation coefficients with human-assigned usefulness scores.}
\label{table:correlations}
\begin{tabular}{lllll}
\hline
\textbf{model} & \textbf{BLEU}  & \textbf{IoU}   & \textbf{$\rho$\textsubscript{BLEU,H}}  & \textbf{$\rho$\textsubscript{IoU,H}}   \\ \hline
natural  & 0.25 & 0.45 & 0.62 & 0.70 \\ 
docstring & 0.18 & 0.40 & 0.57 & 0.65  \\
no comments & 0.06 & 0.27 & 0.73 & 0.74  \\ \hline
all models & 0.16 & 0.37 & 0.63 & 0.69  \\ \hline
\end{tabular}

\end{table}

Table ~\ref{table:correlations} shows the metric results for each model.
$\rho$\textsubscript{BLEU,H} and $\rho$\textsubscript{IoU,H} denote the Pearson correlation between the metrics and human judgments for usefulness, computed using the Pearson product-moment correlation method~\cite{pearson}.
\footnote{To be able to correlate the 4-point reviewer scale with the other metrics, we convert from the 4-point scale to the interval 0-1 as follows: Strongly disagree = 0, Disagree = 1/3, Agree = 2/3, Strongly agree = 1.}
We observe that across all the models in our test, the IoU metric correlates more strongly with human judgements than BLEU.
Furthermore, the correlation factors for IoU are also more consistent across models.

\begin{figure}

\includegraphics[width=0.45\textwidth]{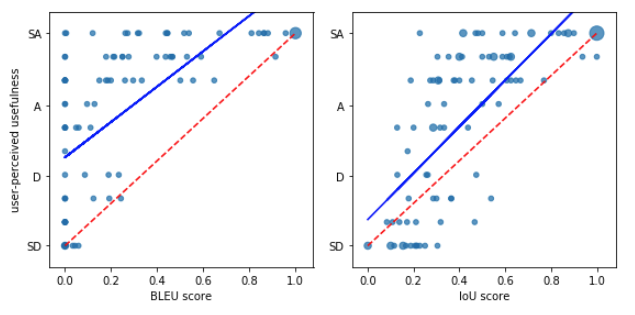}
\caption{Scatter plots that map out the relation between the metric scores (horizontal axis) and the user-perceived usefulness (vertical axis) for the predictions of the \texttt{natural} model.}
\label{fig:scatterplots}
\vspace{-0.5em}

\end{figure}

Figure~\ref{fig:scatterplots} visualizes the relation between the two metrics (BLEU and IoU) and the user-perceived usefulness with the predictions of the \texttt{natural} model.  The best linear fit for the data points is shown in blue, while the red dotted line visualizes the theoretical line on which the metric and usefulness would have perfect correlations.

We manually examined 30 cases where the difference between IoU score and usefulness score was larger than 0.33. Under a perfect correlation, this difference would correspond to one step higher or lower on the annotation scale (e.g. the difference between 'Disagree' and 'Agree').
We found that IoU tends to underestimate the human-assigned usefulness and identified three root causes for the mismatches:

In 12 of 30 failure cases, the prediction contained the target code but also included other code, such as instructions to display or print variables.  While this additional code was often relevant, it significantly decreased the IoU score.

In 7 of 30 failure cases, the model predicted code that satisfied the expressed intent but was syntactically significantly different from the target code. In some cases, the predicted code was more compact and idiomatic than the actual target code. These cases are inherently difficult for any metric that relies purely on the syntactic similarity between the predicted and target code.

In 6 of 30 failure cases, IoU was overly sensitive to a missing import statement in the predicted code, particularly when the code to predict was short, while the annotators seem to care less.

From these observations we conclude that exploring new metrics for both NLGP and code prediction models in general is a relevant area for further research.
\vspace{-0.5em}

\subsection{Summary}

We introduced three code prediction models for Python trained on three different datasets: a model trained on only code (\texttt{no comments}), code with injected docstring comments (\texttt{docstring}) and code with unmodified comments (\texttt{natural}). From an initial benchmark and user study we find that a model trained on natural comments leads to better results in an NLGP context (i.e. predictions based on both prior code and an explicit natural language intent). In future work we want to explore whether we can further boost the prediction quality beyond the \texttt{natural} model by combining natural with injected comments and adapting the docstring comments to better match how developers express intent.

%% file: related_work.tex
\section{Related work}
\label{sec:related_work}

\subsection{Example-Centric Programming}

As mentioned in the introduction, example-centric programming~\cite{brandt10example} tools are a precursor to natural language-guided programming tools. These tools help users more quickly identify code examples from local or online repositories. BluePrint~\cite{brandt10example} allows Adobe Flex users to find relevant code examples from online documentation from within the Adobe Flex Builder IDE. BluePrint takes as input a natural language search query and augments this query with the programming language and framework version used by the developer. Unlike NLGP assistants, Blueprint does not take into account the specific code context and does not adapt the code examples to the specific context of use.

Code assistant tools like Prospector~\cite{mandelin05jungloid} and PARSEWeb~\cite{Thummalapenta07parseweb} focus on the problem of helping developers navigate complex object-oriented APIs.
These approaches share with NLGP the idea of mining common coding idioms from existing code repositories, but do not employ natural language intent to guide the search.

\subsection{Statistical Code Prediction Models}

We cover related work that specifically frames code autocompletion as a statistical code prediction problem.
We divide related work into three categories, depending on what input the prediction model uses: \emph{context-only} models use only the existing code context to predict subsequent code tokens; \emph{intent-only} models use only natural language intent as input without regard for prior code context; finally \emph{context+intent} models use both.

\textbf{Context-Only Models} Tabnine~\cite{Tabnine} and Kite~\cite{Kite} are recent examples of proprietary code autocompletion tools that were trained on code context only.
For both, the aim is to complete the line of code that the developer is actively editing. Tabnine uses a GPT-2 language model~\cite{radford2019language} trained on open source code. A detailed study of GPT-2 autocompletion was carried out by Svyatkovskiy \emph{et al.}~\cite{svyatkovskiy2020intellicode}. They also discussed optimizations such as completion caching to enable efficient deployments of these models. While such statistical code completers can be very effective, they assume that the developer already knows how to start implementing a task. If a developer were to invoke such tools at the end of an inline comment as one would write for an NLGP assistant, these tools would try to autocomplete the comment rather than the next lines of code.

\textbf{Intent-Only Models} Some approaches in this category focus on predicting only API call(s) while others try to predict the entire target code.
Raghothaman \emph{et al.}~\cite{Raghothaman16SWIM} use the clickthrough data of a web search engine to train a model that can translate from user queries into APIs that are likely relevant to the query. They then post-process the relevant APIs into type-safe code examples. This approach does not adapt the generated code to the context of use.
Gu \emph{et al.} \cite{DeepApi} use an RNN encoder-decoder to generate API usage sequences for a given natural language (NL) query. Srinivasan \emph{et al.} \cite{iyer-etal-2018-mapping} predict Java methods given an NL query and a summary of the rest of the class, with a custom LSTM-based encoder-decoder model with attention.
Clement \emph{et al.} \cite{clement2020pymt5} use a T5 encoder-decoder transformer trained on different objectives to predict the Python implementation of a method given its signature and (if available) its docstring. The authors use the Python subset of the CodesearchNet dataset \cite{husain2020codesearchnet} and scrape GitHub repositories for methods with and without docstrings. 
Yin and Neubig \cite{yin2018tranx} generate Python code using an LSTM-based encoder-decoder that produces syntactically correct code by construction.   
Xu \emph{et al.}~\cite{xu2021inide} performed a user study with a code prediction plugin based on an ensemble of TRANX~\cite{yin2018tranx} and a code search model.  The plugin suggests code snippets based on a natural language query that is issued within an IDE. Their setup is therefore closely related to natural language-guided programming, except that the plugin does not leverage the surrounding code. The user study did not provide conclusive evidence that the plugin had a significant influence on programmer productivity: neither on the speed with which programmers solved tasks nor on the correctness of the implementations. 

\textbf{Context+Intent Models} Murali \emph{et al.}~\cite{DBLP:journals/corr/MuraliCJ17a} predict a Java method body given 1) NL keywords and 2) API calls or classes that should be used.  Based on these inputs a probabilistic encoder-decoder named ``Gaussian Encoder-Decoder'' (GED) was used to learn a distribution over simplified control-flow graphs (``program sketches'').
Agashe \emph{et al.}~\cite{agashe-etal-2019-juice} study both API call prediction and full code prediction 
based on the preceding code and a natural language query. They experimented with LSTMs and small Transformer models without pretraining on a large text corpus. Orlanski and Gittens~\cite{orlanski-gittens-2021-reading} studied code generation from StackOverflow questions, which often embed code snippets to provide additional context.

Chen \emph{et al.}~\cite{chen2021evaluating} introduce Codex, a series of large language models (up to 12B parameters) based on the GPT-3~\cite{Brown2020} architecture. By training the models on a large corpus (159GB) of open source code they find that Codex performs significantly better than GPT-3 on the task of predicting full code solutions to programming problems from natural language docstrings. Whereas our work on NLGP focuses on generating small snippets of code to help a programmer more effectively explore and use known APIs, Codex is evaluated on generating functionally correct code. Testing functional correctness of code requires executable unit tests which may not always be available in practical settings.

\subsection{Code Prediction Benchmark Data}

In our search for usable benchmarks, we find that existing benchmarks typically consist of intent and target code while benchmarks suitable to test NLGP assistants require context, intent, and target code.

\textbf{Intent/Target Benchmarks:}
Yin \emph{et al.}~\cite{10.1145/3196398.3196408} create a curated dataset called CoNaLa \cite{Conala} from Stack Overflow posts. 
Heyman \emph{et al.} \cite{heyman2020neural} created a benchmark for ``annotated code search'': the retrieval of code snippets (target code) annotated with a short natural language description (intent).   
Yao \emph{et al.} \cite{DBLP:journals/corr/abs-1803-09371} mined question-code (Intent/target) pairs in Python and SQL.
Hamel Husain \emph{et al.}~\cite{husain2020codesearchnet} collected query/code (intent/target) pairs from Bing search queries that have high click-through rates to code written in Go, Java, JS, PHP Python or Ruby. 
Barone \emph{et al.}~\cite{DBLP:journals/corr/BaroneS17} extracted 100K target/intent samples from Github projects.  
The intent is retrieved from docstrings that describe function declarations and bodies.
Chen \emph{et al.}~\cite{chen2021evaluating} introduce \emph{HumanEval}, a benchmark of 164 manually composed programming problems and their Python solutions, consisting of a function signature, docstring, function body and several unit tests. While this benchmark is useful to measure functional correctness of generated code, the benchmark problems are typical programming challenges focused on mathematical concepts using built-in abstractions like numbers and lists, and are therefore not suitable to assess code generation for programmer intents in API-rich settings such as the Python data science domain. The problems are also self-contained, not requiring prior code context.

\textbf{Context/Intent/Target Benchmarks:}
Agashe \emph{et al.}~\cite{agashe-etal-2019-juice} created a dataset of 3.7K curated examples called JuICe. 
The samples are extracted from Jupyter notebooks containing Python code.  
As these notebooks were originally created as student assignments, the natural language intents tend to be long, descriptive and often contain information that is only loosely related to the target code.
An average intent in the dataset measures 58.33 tokens and is therefore less suited for NLGP, where we expect the intent to be formulated as a short query of between 3 and 10 tokens. This expectation is based in part on observations from a user study conducted by Xu \emph{et al.}~\cite{xu2021inide}.

\subsection{Program Synthesis}

Program synthesis methods~\cite{gulwani17program} study the broader problem of generating programs from specifications. Specifications can range from highly formal and unambiguous (e.g. a formula in logic) to informal and ambiguous (e.g. input-output examples, natural language or program sketches~\cite{solarlezama08program}). Most closely related to NLGP is the idea of program synthesis from natural language input~\cite{desai16program}. These methods focus on translating a natural language intent (often just a single sentence) into a short program that covers the intent. A key difference with NLGP is that these methods typically focus on helping end-users in specific domains: the natural language input is restricted to a specific application domain and the programs are written in a domain-specific language (DSL) that is often custom-built to solve a specific problem. This contrasts with NLGP which is aimed at helping professional software developers solve a variety of tasks using a general-purpose programming language.

%% file: research_agenda.tex
\section{A Research Agenda for Natural Language-Guided Programming}
\label{sec:research_agenda}

As with any proposal that aims to offer radically new ways to program computers, the idea
of writing code guided by free-form natural language brings with it a whole new range of problems
and unexplored areas.
What research questions does the programming community need to address to turn
natural language-guided programming from a research idea into a reliable ``proven'' method of
programming? We list significant open questions that remain unanswered by the case study
presented in this work. Each of these represents a major avenue for future research in
natural language-guided programming:

    \smallskip\noindent\textbf{More Diverse Training Data}
    It is to be expected that training models on more source code
    will further increase the quality of code predictions. In addition, rather than simply training models on more code, it would be useful to consider additional sources of NL intent/code pairs, such as tutorial documentation or Q\&A forum threads (such as those found on Stack Overflow).
    
    \smallskip\noindent\textbf{Better Benchmark Datasets}
    Progress in machine learning and NLP is often driven by high-quality benchmarks (e.g., the GLUE benchmark~\cite{wang2018glue}). In the same vein, we believe better benchmarks for code prediction are a key enabler for better NLGP assistants. In this work we have taken the first steps towards this goal, but our benchmark remains limited in size (201 examples) and in scope (Python data science). We hope that the community will advance these efforts.
   
    \smallskip\noindent\textbf{Better Metrics}
    Development of a benchmark not only entails creating curated triplets of realistic code contexts,
    NL intents and ground-truth code completions, but also entails finding better code
    scoring metrics whose output correlates even better with user-perceived usefulness.

    Right now, the most effective way to measure the usefulness of a code-prediction tool is to have human experts rate the predicted code in relation to the
    stated task and the given code context. This method is not very scalable, especially
    when considering comparing multiple (or multiple versions of) code prediction models.

    What is needed is an easy to calculate and objective metric that can score the output
    of code prediction models with reference to one or more ground-truth solutions.

    In this work, we used standard metrics such as BLEU and IoU to compute
    the similarity between predicted code and the ground-truth target code.
    We have shown that these metrics correlate with user-perceived usefulness to some extent (Section~\ref{sec:metrics}). There is ample opportunity to improve upon these metrics with new metrics more specifically tailored to code.
    
    \smallskip\noindent\textbf{Effect on Productivity}
    Does NLGP positively affect developer productivity as measured by e.g. the time to complete
    set programming tasks?
    Even though the ultimate goal of code-prediction models is to maximize the productivity 
    of developers and the quality of the code, there is a relative paucity of research that
    quantifies these claims. Recent work by Xu \emph{et al.}~\cite{xu2021inide} aims to address this through
    a controlled user study where two groups of programmers were tasked to complete a set of
    well-defined programming tasks, with and without the help of a code-prediction tool.
    The results from the study were inconclusive as to the positive effect of the code-prediction
    tool under study. There is a clear need for more of these studies with larger participation,
    more diverse tasks and more code-prediction tools.

    \smallskip\noindent\textbf{Effect on Code Quality} Does NLGP positively affect the quality of code as measured by e.g. reported bugs
    attributed to code (partially) suggested by NLGP assistants? Does NLGP positively affect the
    maintainability of code?
    
    \smallskip\noindent\textbf{Effect on Learning Curve} What is the effect of NLGP on the learning curve of a developer? For example,
    are NLGP assistants better suited to junior developers or are they helpful across many levels
    of prior coding experience? Are NLGP assistants more useful for developers new to a project or
    do they remain useful even for senior developers on the team?
    
    \smallskip\noindent\textbf{Impact of Text-to-Code Ratio} How does the effectiveness of NLGP relate to the ratio of code versus natural language
    text in coding environments? Our case study focused on Jupyter notebooks where the ratio
    of natural language text (in-line comments, markdown text cells) compared to code is likely
    higher than in a typical Python script (a `.py` source file). It is intuitively clear
    that a higher ratio of text-to-code will help NLGP, but we have yet to establish objective
    relationships between text-to-code ratio and NLGP effectiveness as measured through
    benchmarks.

   \smallskip\noindent\textbf{Inference Latency} For modern neural architectures such as Transformers, deep learning researchers have observed that larger models perform better. 
   Our initial experience in training GPT-2 models of various sizes (not further detailed in this paper) for the NLGP task confirms this observation. We have deliberately kept the model size constrained to keep the latency of code predictions within an acceptable threshold of 2 seconds. We expect increasing the inference speed of language models by algorithmic or hardware improvements will be an important driver to enable larger and therefore better code predictions.
   
   \smallskip\noindent\textbf{Effective Use of Code Context} How to leverage the code context more effectively?
   Because for transformer architectures such as GPT-2, memory and time complexity scale quadratically in the sequence length, it is infeasible to provide an entire code file as context to such models.
   Our case study uncovered that a significant proportion of the mistakes occurred when relevant code (e.g. import statements) fell out of the model's context window. For example, we observed that the language models we trained at times predict functional calls that look relevant but do not exist. We conjecture that this phenomenon is caused by training with a limited context window, because during training the model will at times be forced to predict function calls without seeing its definition or imports.
   Therefore, new strategies to select what parts of a code file will be included in the context window, more efficient tokenization methods (i.e. encoding the same code with fewer tokens), and exploring linear/subquadratic transformer variants could all lead to more informed predictions. 

   \smallskip\noindent\textbf{API Versioning} Building an NLGP assistant by training a model on existing code runs the risk of biasing the model's predictions towards older or more frequently used versions of common APIs.
   Ideally, an NLGP tool would also have access to the precise versions of the libraries used by the developer so that it can tailor its code suggestions to those versions. This would help overcome a key limitation of example-centric programming, as studies of Stack Overflow found that code found in answers to coding questions was frequently obsolete~\cite{Ragkhitwetsagul21toxic}, with one study finding that
   for a sample of known-obsolete answers only 20.5\% were updated to reflect the latest API usage~\cite{zhang21empirical}.

    \smallskip\noindent\textbf{Impact of Interactive Programming} Can interactive programming environments
    further improve the effectiveness of NLGP by giving the NLGP assistant access to
    (a description of) the runtime values manipulated by the code?

    Interactive programming environments such as notebook environments
    (Jupyter, Zeppelin, Observable, etc.) or IDEs that prominently support
    read-eval-print loops (e.g. DrRacket, BlueJ) offer the capability to execute
    small code fragments and get immediate feedback on their runtime effects.
    For example, in a Jupyter notebook, the output of a code cell is often
    inserted as rich output in the notebook environment itself (as a graphic, a
    structured table or as plain text).
    
    Taking this one step further, ``Live Programming'' environments~\cite{mcdirmid13usable}
    aim to merge code and runtime context even further, giving near-continuous feedback
    on the runtime values stored in program variables. We conjecture that an NLGP assistant
    could make effective use of these additional context inputs to improve its suggestions.


%% file: conclusion.tex
\section{Conclusion}

We define natural language-guided programming as the programming practice of using intelligent code completion tools to automate routine programming tasks by stating the desired intent using natural language. An NLGP assistant is a tool that autocompletes a piece of code guided by natural language intent.

We demonstrate natural language-guided programming for automating routine data science and machine learning tasks using Python. We contribute the design, implementation and evaluation of a proof-of-concept NLGP assistant based on language modeling.

We conduct experiments with pretrained models (GPT-2), revealing that preparation of the data to contain a good mix of natural language intent and code is critical to improve code prediction quality. Our experiments suggest that comments that occur naturally in the code are sufficient for language models to learn the relationship between intent and code. Our docstring injection method further indicates that NLGP can be made feasible in domains where the source code lacks good inline comments.

We construct a curated benchmark to measure the quality of code predictions. 
Our initial human evaluation study provides evidence that our best models can generate code predictions that expert data scientists find useful, and that are compatible with the context of use. As such, our work can be seen as a first step towards making automatic example embedding a reality.

Much work remains to be done to turn NLGP from an initial idea into a practical, reliable programming practice. We end the paper with a Research Agenda for NLGP, inviting the programming research community to work on better benchmarks, to set up user studies to quantify the impact on productivity, and to invent novel metrics to automate the scoring of code predictions. Our initial experiments reveal inconsistencies between widely used metrics and human judgments, which we hope will inspire others to invent better alternatives.

%% file: appendix.tex
\appendix
\section{Appendix}

\subsection{Annotation Process}\label{ap:annotation-process}
\subsubsection{Annotation Guidelines}\label{ap:annotation_guidelines}
The annotators received the following guidelines:

\begin{itemize}[leftmargin=*]
    \item Skip candidates for which the candidate intent:
    \begin{itemize}
    \item only contains commented source code
    \item does not express the intent of (part of) the candidate target code (e.g. skip comments from exercise notebooks such as \verb|"Start your code here"| )
    \item is domain-specific and cannot be translated into the target code without expert knowledge about a particular domain. The intent should be expressed in terms that refer to the functionality of Python/ Python libraries, it should not express the more high-level, domain-specific goal for why the libraries are needed. Note that this guideline does not imply that the intent has to include library names or API calls.
    \end{itemize}

	\item Skip candidates for which the target code:
	\begin{itemize}
 	\item does not contain at least one non-trivial API call (e.g. \verb|"Fg = Fn * g"|);
    \item exclusively consists of setup/initialization code;
    \item is a non-idiomatic implementation of the intent
    \end{itemize}

    \item For the remaining test cases:
    \begin{itemize} 
    \item select the target code $t$ from the candidate target code $t_c$
    \item reformulate the candidate intent $i_c$ to make it more realistic, if necessary. For example, "and now plot the data" can be reformulated as "plot the data". 
    However, to avoid that the intent would be systematically biased towards the annotator preferences, 
    annotators are not allowed to further reformulate the original intent. 
    Similarly, annotators are instructed to not correct potential typos.
    \end{itemize}
\end{itemize}

\begin{figure}
\includegraphics[width=0.45\textwidth]{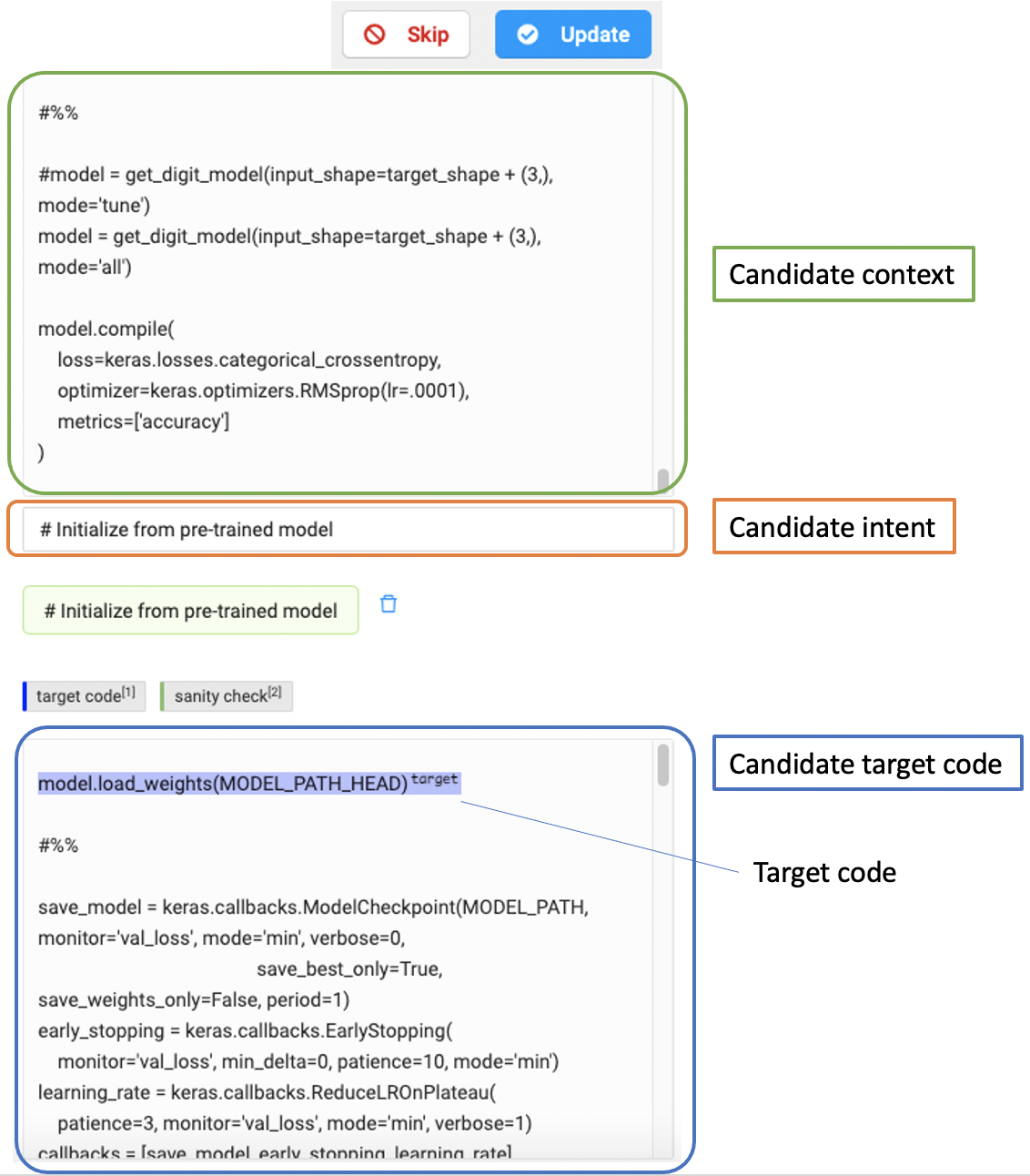}
\caption{Screenshot of the user interface for annotating candidate test cases to create the NLGP benchmark.}
\label{fig:user-interface}
\end{figure}

\subsubsection{Annotation Interface}
Figure~\ref{fig:user-interface} shows the user interface for annotating candidate test cases to create the NLGP benchmark. It is implemented as a web service using the Label Studio framework.\footnote{\url{https://labelstud.io/}}

\subsubsection{Inter-Annotator Agreement}
\sloppy We evaluated the inter-annotator agreement on three aspects: the decision to accept or skip a test case, the intent selection, and the target code selection. The Fleiss kappa score with regard to accepting/skipping test cases is 0.515. This reflects moderate agreement~\cite{viera2005understanding}. Out of the candidate test cases that were accepted by at least two annotators, annotators had the same intent for 74\% of the cases and had annotated the same target code for 71\% of the cases. For the cases without exact agreement, the intents had an average edit distance of 3.3 tokens and the target code snippets differed with an average of 3.2 lines of code.

\lstset{basicstyle=\footnotesize, keepspaces=true, columns=flexible}

\begin{figure*}
\includegraphics[width=0.7\textwidth]{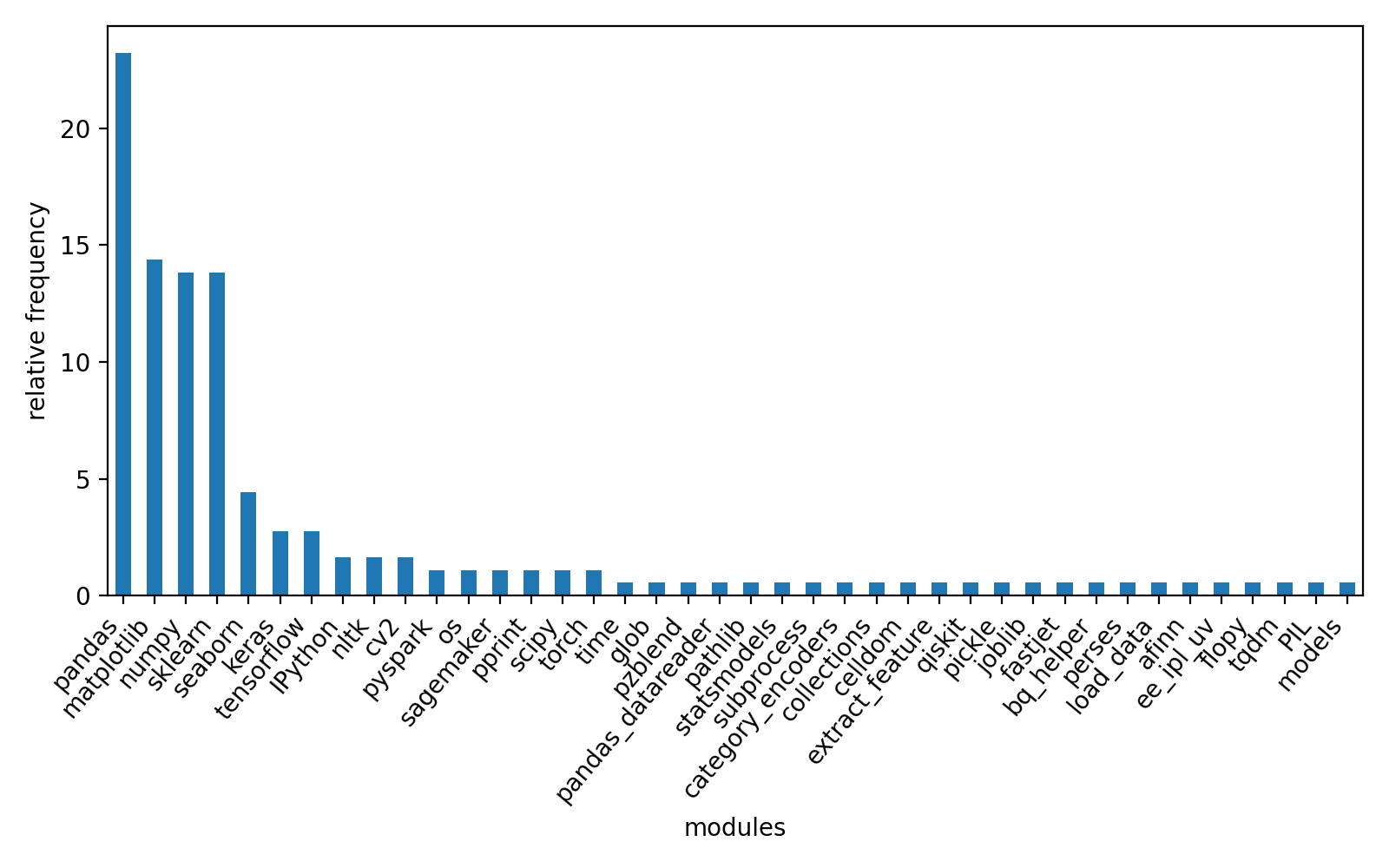}
\vspace{-1em}
\caption{Plot of the frequencies with which modules are used in the target code of the 201 test cases in the NLGP benchmark. }
\label{fig:modules_target_code}
\end{figure*}

\begin{figure*}
\includegraphics[scale=0.23]{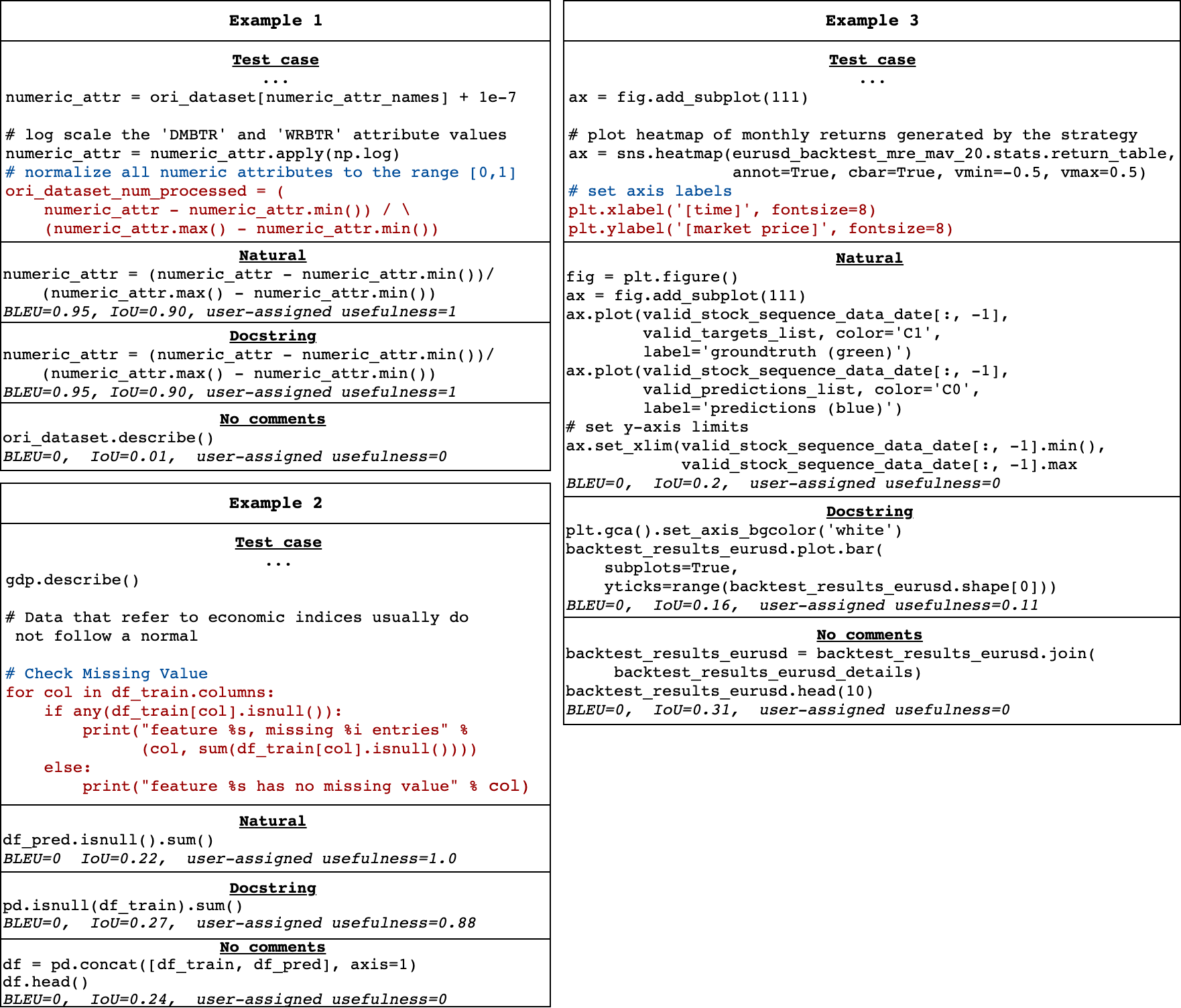}
\caption{Three test cases from the NLGP benchmark and the predictions for the natural, docstring and no comments models. We display each test case with the truncated context in black, the intent in blue and the target code in red.}
\label{fig:lgp-examples}
\end{figure*}

\subsection{NLGP Benchmark Library Distribution}
Figure~\ref{fig:modules_target_code} gives insight into what libraries are used in the target code fragments in the NLGP benchmark. For each of the 201 test cases, we analyzed the imports and calls in the target code and attempt to resolve these to their root module. We rely on a similar resolution method to what was used for the docstring injection (see \ref{sec:data_augmentation}). Note that there will be cases where the root module is not resolved correctly, but overall the method should be accurate enough to capture the module distribution.

\subsection{Examples}\label{ap:examples}

In Figure \ref{fig:lgp-examples}, we list three test cases from the NLGP benchmark and the predictions made by the three models under test.   We also provide the scores assigned by BLEU and IoU as well as the average usefulness score assigned by the users. Note that due to space constraints we truncated the context code.